\documentclass[preprint,floatfix,11pt] {revtex4}

\usepackage{graphicx}
\usepackage{subfigure}
\usepackage{xcolor}
\usepackage{amsmath}
\usepackage{enumerate}
\usepackage{tabularx}
\usepackage{amsfonts}
\usepackage{amssymb}
\usepackage{color}
\definecolor{kugreen}{RGB}{5,93,0}
\definecolor{kugreenlys}{RGB}{132,158,139}
\definecolor{kugreenlyslys}{RGB}{173,190,177}
\definecolor{kugreenlyslyslys}{RGB}{214,223,216}
\definecolor{darkblue}{rgb}{0,0,0.5}

\newcommand{\bd}{\begin{document}}
	\newcommand{\ed}{\end{document}}
\newcommand{\bc}{\begin{center}}
	\newcommand{\ec}{\end{center}}
\newcommand{\bfr}{\begin{flushright}}
	\newcommand{\efr}{\end{flushright}}
\newcommand{\lt}{\left}
\newcommand{\rt}{\right}
\newcommand{\vs}{\vspace}
\newcommand{\hs}{\hspace}
\newcommand{\beq}{\begin{equation}}
\newcommand{\eeq}{\end{equation}}
\newcommand{\lb}{\linebreak}
\newcommand{\pb}{\pagebreak}
\newcommand{\mb}{\makebox}
\newcommand{\fb}{\framebox}
\newcommand{\mc}{\multicolumn}
\newcommand{\ben}{\begin{enumerate}}
	\newcommand{\een}{\end{enumerate}}
\newcommand{\bit}{\begin{itemize}}
	\newcommand{\eit}{\end{itemize}}
\newcommand{\oln}{\overline}
\newcommand{\un}{\underline}
\newcommand{\lefq}{\lefteqn}
\newcommand{\ba}{\begin{array}}
	\newcommand{\ea}{\end{array}}
\newcommand{\beqa}{\begin{eqnarray}}
\newcommand{\eeqa}{\end{eqnarray}}
\newcommand{\beqas}{\begin{eqnarray*}}
	\newcommand{\eeqas}{\end{eqnarray*}}
\newcommand{\bfg}{\begin{figure}}
	\newcommand{\efg}{\end{figure}}
\newcommand{\bds}{\begin{displaymath}}
\newcommand{\eds}{\end{displaymath}}
\newcommand{\btb}{\begin{tabbing}}
	\newcommand{\etb}{\end{tabbing}}
\newcommand{\para}{\parallel}
\newcommand{\pad}{\partial}
\newcommand{\nn}{\nonumber}
\newcommand{\la}{\leftarrow}
\newcommand{\ra}{\rightarrow}
\newcommand{\lgla}{\longleftarrow}
\newcommand{\lgra}{\longrightarrow}
\newcommand{\La}{\Leftarrow}\newcommand{\Ra}{\Rightarrow}
\newcommand{\Lra}{\Leftrightarrow}
\newcommand{\Lgla}{\Longleftarrow}
\newcommand{\Lgra}{\Longrightarrow}
\newcommand{\lan}{\langle}
\newcommand{\ran}{\rangle}
\renewcommand{\a}{\alpha}
\renewcommand{\b}{\beta}
\newcommand{\g}{\gamma}
\newcommand{\G}{\Gamma}
\renewcommand{\d}{\delta}
\newcommand{\eps}{\epsilon}
\newcommand{\Th}{\Theta}
\newcommand{\s}{\sigma}
\newcommand{\lam}{\lambda}
\newcommand{\D}{\Delta}
\newcommand{\ds}{\displaystyle}
\newcommand{\vare}{E}
\newcommand{\pr}{\prime}
\newcommand{\ro}{\rho}
\newcommand{\nab}{\nabla}
\newcommand{\m}{\mu}
\newcommand{\n}{\nu}
\newcommand{\Sg}{\Sigma}
\newcommand{\p}{\pi}
\newcommand{\R}{I\!\!R}
\newcommand{\om}{\omega}
\newcommand{\Om}{\Omega}
\newcommand{\ovra}{\overrightarrow}
\newcommand{\ze}{\zeta}
\newcommand{\vart}{\vartheta}
\newcommand{\tri}{\triangle}
\newcommand{\f}{\frac}
\newcommand{\iny}{\infty}
\newcommand{\pro}{\propto}
\renewcommand{\arraystretch}{1.25}
\begin{document}

\title{Analytical solution of {\color{red} $D$ dimensional Schr\"odinger equation for Eckart potential} with a new improved approximation in centrifugal term}

\author{Debraj Nath}
\altaffiliation{Corresponding author. Email: debrajn@gmail.com} 
\affiliation{Department of Mathematics, Vivekananda College, Kolkata-700063, WB, India} 

\author{Amlan K.~Roy}
\altaffiliation{Corresponding author. Email: akroy@iiserkol.ac.in, akroy6k@gmail.com.} 
\affiliation{Department of Chemical Sciences Indian Institute of Science Education and Research (IISER) Kolkata, Mohanpur-741246, Nadia, WB, India}

\begin{abstract}
Analytical solutions are presented for eigenvalues, eigenfunctions of {\color{red} D-dimensional Schrodinger equation having Eckart potential} within 
Nikiforov-Uvarov method. This uses a new, improved approximation for centrifugal term, from a combination of Greene-Aldrich and Pekeris 
approximations. Solutions are obtained in terms of hypergeometric functions. It facilitates an accurate representation in entire domain. 
Its validity is illustrated for energies in an arbitrary $\ell \neq 0$ quantum state. Results are compared for a 
chosen set of potential parameters in different dimensions. In short, a simple accurate 
approximation is offered for Eckart and other potentials in quantum mechanics, in higher dimension.

 
\vspace{10mm}

\end{abstract}
\maketitle
\section{Introduction}
The exponential diatomic molecular Eckart potential \cite{eckart1930},  
\begin{equation}\label{Eckart}
v_e(r)=-\frac{\a e^{-r/a}}{1-e^{-r/a}}+\f{\b e^{-r/a}}{(1-e^{-r/a})^2},~\a,\b>0. 
\end{equation}
is used as an important model in chemical physics \cite{cooper1995}. The depth is controlled by two positive real parameters $\alpha$ and $\beta$, while the range is governed by a positive parameter, $\alpha$, having the dimension of inverse length. It is related to the Schi\"oberg potential by a change of parameter. It has a minimum value of $- \f{(\a-\b)^2}{4\b}$ at $r= a \ \ln \f{(\a+\b)}{(\a - \b)},$ for $\a > \b$. Like the other few familiar exponential-type model potentials such as, Morse, Woods-Saxon, Hulth\'en, Manning-Rosen, the relevant Schr\"odinger equation (SE) in this case too offers \emph{exact} analytical solution \emph{for $\ell =0$}, but not for $\ell \ne 0$ states. {\color{red} However, a number of elegant approximations 
have been put forth for non-zero $\ell$ states with varied success.} It may be noted that the Eckart potential and its PT-symmetric version can be considered as special cases of the multi-parameter exponential type potential \cite{jia2003}.  

Analytical bound state energy expressions and radial wave functions for arbitrary $\ell$ states are derived by using various approximations to the centrifugal term \cite{dong2007,pekeris.ap3, taskin2010,stanek2011} in amalgamation with a host of techniques, such as, supersymmetric shape invariance approach \cite{diao2009,onate2018} combined with a functional analysis \cite{diao2009}, asymptotic iteration \cite{falaye2012}, tridiagonal representation \cite{zhang2012}, Feynman path integral approach \cite{diaf2015}, etc. An approximate solution of the SE with Eckart potential and its parity-time symmetric version was investigated \cite{zhang2009} using a Nikiforov-Uvarov (NU) method. The same for scattering states were expressed in terms of the generalized hypergeometric functions $_2F_1(a,b;c;z)$, and phase shifts were analyzed \cite{chen2008, wei2008}. The relativistic bound- \cite{wei2009, liu2009} and scattering-state \cite{wei2009} solutions of Klein-Gordon equation have been reported. Apart from energies, information theoretical studies were pursued, \emph{viz.}, Shannon entropy \cite{pooja2016}, R\'enyi entropy \cite{onate2018}, etc. Variants of this potential, like deformed hyperbolic Eckart \cite{mincang2013}, Eckart-like \cite{pekeris.ap3}, Hua plus modified Eckart \cite{onyenegecha2020}, generalized Deng-Fan plus deformed Eckart \cite{awoga2013} potential have been considered as well. The thermodynamic properties of anharmonic Eckart potential have also been the subject of investigation \cite{ortega2017}.  
The above discussion suggests that, a reasonable number of good-quality methods are available for low-lying (especially, $\ell =0$) states of this potential. But such attempts are rather limited for high-lying states belonging to the non-trivial $\ell \neq 0$ case. Also it is worth mentioning that a majority of the articles are devoted to eigenvalues and eigenfunctions, mainly in 3D. A review of the available methods further indicates that, a decent number of works resort to the approximation of centrifugal term from physical/mathematical consideration. Herein lies the objectives of this work, which is two-fold. At first, we would like to make an analysis and test performance of a recently developed, simple approximation for centrifugal term in the context of Eckart potential. So far, this has been applied to shifted Deng-Fan, Manning-Rosen and P\"oschl-Teller molecular potentials \cite{nath2021, nath2021.epjp}. This will extend the scope of applicability to a wider range of physical systems. Another objective is to report the results in D dimension, for which the references are very rare \cite{Hulthen.greene15}. 

The article is organized as follows: Sec.~ \ref{Sec2.Exac} gives an overview of various approximations to the centrifugal term, including the newly proposed one \cite{nath2021, nath2021.epjp}, for solving the SE for Eckart potential, within the NU method. The resultant analytic expressions are derived for radial wave functions, eigenvalues, normalization constants. 
Wherever possible, the computed energies in Sec.~\ref{Sec3.results} are discussed and compared with available literature results. Finally, a few remarks are made in Sec.~\ref{Sec4.con}, along with the future prospect of our scheme.  

\section{Theoretical analysis}\label{Sec2.Exac}
Let us consider the following SE,  
\beq\label{Schro.Eq}
-\f{\hbar^2}{2\bar{\mu}}{\bf\nabla}_D^2\psi+V({\bf r})\psi({\bf r})=E\psi({\bf r}),
\eeq 
where $\overrightarrow{\bf\nabla}$ is $D$-dimensional gradient operator, ${\bf r}=(x_1,x_2,...,x_D)=(r,\theta_1,\theta_2,...,\theta_{D-1})=(r,\Om_{D-1})$, $r=|{\bf r}|=\sqrt{\sum\limits_{i=1}^Dx_i^2}$, $x_i=r\cos\theta_i\prod\limits_{k=1}^{i-1}\sin\theta_k$, $0\le\theta_i<\pi,~i=1,2,...,D-2$, $0\le \theta_{D-1}<2\pi$, while, 
\beq
{\bf\nabla}_D=\left(\f{\partial}{\partial r},\f{1}{r}\f{\partial}{\partial\theta_1},\f{1}{r\sin\theta_1}\f{\partial}{\partial\theta_2},...,\f{1}{r\prod_{j=1}^{k-1}\sin\theta_{j}}\f{\partial}{\partial\theta_k},...,\f{1}{r\prod_{j=1}^{D-2}\sin\theta_{j}}\f{\partial}{\partial\theta_{D-1}}\right),
\eeq 
is the $D$-dimensional gradient operator. Then the Laplacian operator ${\bf\nabla}_D^2$ is defined by,
\beq
{\bf \nabla}_D^2=\f{1}{r^{D-1}}\f{\partial}{\partial r}\left(r^{D-1}\f{\partial}{\partial r}\right)-\f{\Lambda_D^2}{r^2},
\eeq
where $\Lambda_D$ is the $D$-dimensional generalized angular momentum operator, given by,
\beq
\ds\Lambda_D^2=-\sum\limits_{i=1}^{D-1}\f{(\sin\theta_i)^{i+1-D}}{(\prod_{j=1}^{i-1}\sin\theta_{j})^2}\f{\partial}{\partial\theta_i}\left((\sin\theta_i)^{D-i-1}\f{\partial}{\partial\theta_i}\right),
\eeq 
$\bar{\mu}$ is reduced mass of diatomic molecule, and $r$ is the internuclear distance. Let
\beq
\psi({\bf r})=r^{-\f{D-1}{2}}R(r)Y_{\ell,\{\mu\}}(\Om),
\eeq 
be the solution of Eq.~(\ref{Schro.Eq}). Then it is obtained that, 
\beq\label{Eq.R}
\f{d^2R}{dr^2}+\left[\f{2\bar{\mu}}{\hbar^2}E-\f{2\bar{\mu} }{\hbar^2}V(r)-\f{L(L+1)}{r^2}\right]R=0,
\eeq 
and
\beq
\ds\Lambda_D^2 Y_{\ell,\{\mu\}}=\ell(\ell+D-2) Y_{\ell,\{\mu\}},
\eeq 
where $\ell(\ell+D-2)$ is the separation constant, and $L=\ell+\f{D-3}{2}$. Then following \cite{dehesa.jmp.2007,Srivastava-Daoust}, 
\beq\label{Ylm}
Y_{\ell,\{\mu\}}=N_{\ell,\{\mu\}}e^{im\theta_{D-1}}\prod\limits_{j=1}^{D-2}C_{\mu_j-\mu_{j+1}}^{\a_j+\mu_j+1}(\cos\theta_j)(\sin\theta_j)^{\mu_j+1},
\eeq 
where $C_{i}^{j}(t)$ is the Gegenbauer polynomial in $t$ of degree $i$ with parameter $j$, whereas 
\beq
N_{\ell,\{\mu\}}=\f{1}{2\pi}\prod\limits_{j=1}^{D-2}\f{(\a_j+\mu_j)(\mu_j-\mu_{j+1})!\G(\a_j+\mu_{j+1})^2}{\pi 2^{1-2\a_j-2\mu_{j+1}}\G(2\a_j+\mu_j+\mu_{j+1})},
\eeq 
is the normalization constant. Here $(\ell,\{\mu\})=(\mu_1,\mu_2,...,\mu_{D-1})$, $\ell=\mu_1\ge\mu_2\ge...\ge\mu_{D-2}\ge|\mu_{D-1}|=|m|$, $\ell=0,1,2,...$, $m=0,\pm1,\pm2,...$, $\a_j=(D-j-1)/2$, $D=3,4,5,6,\cdots$.
To find the general solution of resulting radial SE of Eq.~(\ref{Eq.R}), we will use a transformation, $s=e^{-r/a}$, 
and a set of functions, $\left\{1,\f{s}{1-s},\f{s^2}{(1-s)^2}\right\}$, for which the centrifugal term is approximated in different forms, 
such as Greene-Aldrich \cite{greene}, Pekeris-type \cite{Pekeris}. Recently, the authors have proposed an intuitive combination of these 
two approximations \cite{nath2021,nath2021.epjp} with considerable success. 

\subsection{Approximation to the centrifugal term}
\begin{figure} 
	\centering
\hspace{-2cm}\includegraphics[width=20cm,height=14cm]{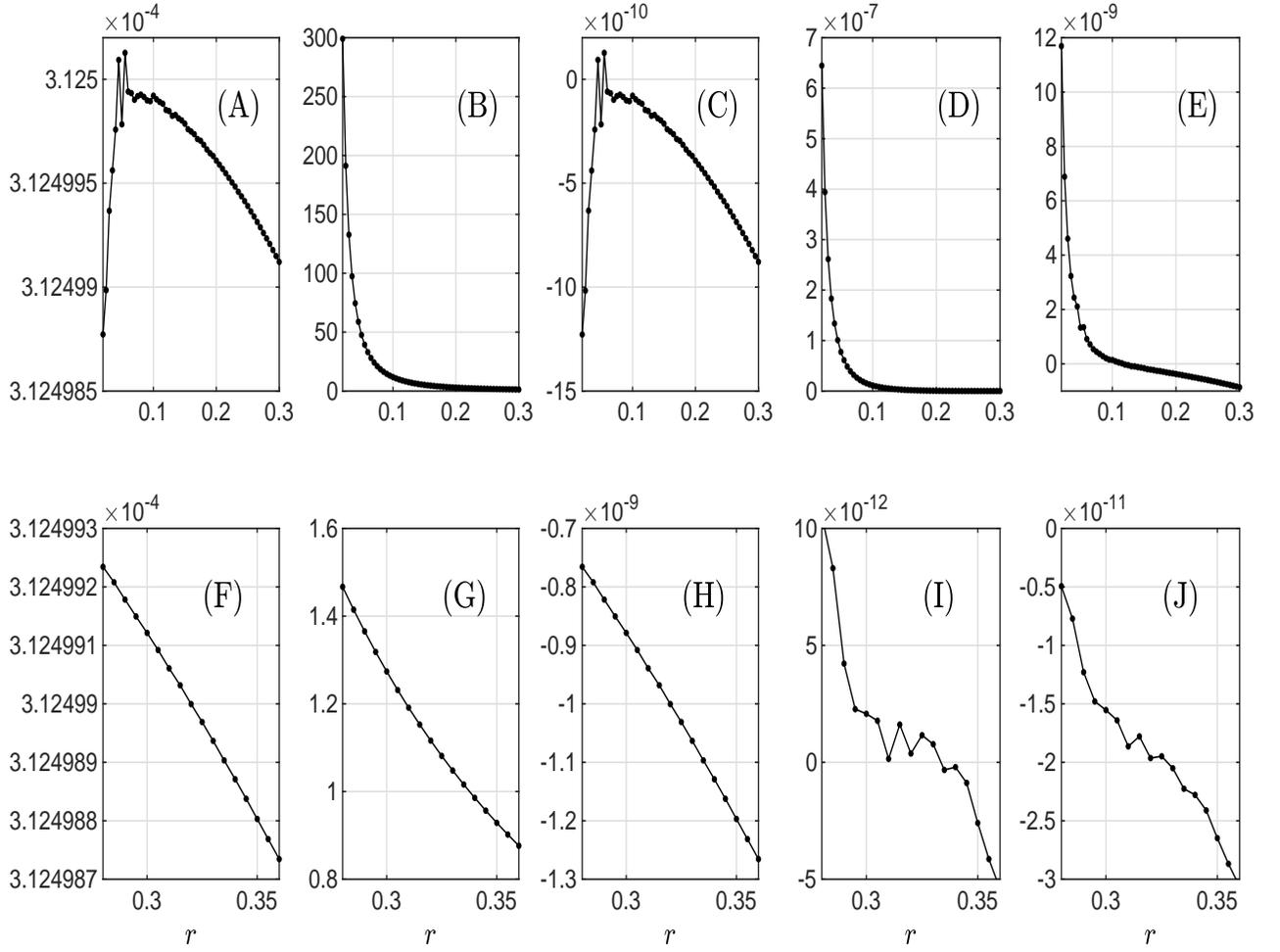}
	\caption{\label{Fig1.approx} Plot of the difference, $\ell(\ell+1)\left(\f{1}{r^2}-f_i(r)\right)$, for $i=1,2,3,4$. In (A), (F) $i=1$; (B), (G) $i=2$; (C), (H) $i=3$; (D), (I) $i=4$. (E) and (J) refer to $f_5^{(a)}$ and $f_5^{(b)}$ respectively. The parameters are: $\ell=2$, $(\hbar,\bar{\mu})=(1,1)$, 
$(a,\a,\b)=(1/0.025,1/a,0.0001)$, $(\xi_1,\xi_2)=(1.1,0.98)$, $D=3$. See text for details.}
\end{figure}

Let us consider the following relation:
\beq\label{greene.app1}
\f{1}{r^2}\approx f_i(r)=\f{1}{a^2}\left(x_{1i}+\f{x_{2i}s}{1-s}+\f{x_{3i}s^2}{(1-s)^2}\right),s=e^{-r/a},~i=1,2,3.
\eeq 
The approximation $f_1(r)$, with $x_{11}= 0,x_{21}=x_{31}=1$, is commonly used in references \cite{dong2007,zhang2012,wei2008,greene}, while for $x_{12}=0, x_{22}=\xi_1, x_{32}=\xi_2$ and $\xi_1,\xi_2$ as two adjustable dimensionless parameters, the 
approximation $f_2(r)$ is considered by the authors of \cite{taskin2010,onate2018}. Apart from that, the approximation 
$f_3(r)$ is employed in reference \cite{f3,f3.2,f3.3,f3.4,f3.5}, having $x_{13}=\f{1}{12},x_{23}=x_{33}=1$. The approximations $f_1,f_2,f_3$ are good 
near the origin \cite{dong2007,taskin2010,onate2018,zhang2012,wei2008,greene}. Moreover, the centrifugal term 
$\f{1}{r^2}$ is approximated near the point $r=r_0$, as below, following the references 
\cite{pekeris.ap3,Pekeris,pekeris.badawi,pekeris.ap4},  
\beq\label{pekeris.app1}
\f{1}{r^2}\approx f_4(r)=\f{1}{a^2}\left(x_{14}+\f{x_{24}s}{1-s}+\f{x_{34}s^2}{(1-s)^2}\right),
\eeq
where
\beq
\ba{l}
x_{14}=\f{1}{u^4}\left[(3+u)s_0^2+(2u-6)s_0+(3-3u+u^2)\right],\\
x_{24}=\f{2}{u^4}(1-s_0)^2\left[3+u+\f{2u-3}{s_0}\right],\\
x_{34}=-\f{1}{u^4}(1-s_0)^3\left[\f{3+u}{s_0}+\f{u-3}{s_0^2}\right],\\
s_0=e^{-u},u=r_0/a.
\ea 
\eeq
Here $r_0$ is a positive real number, having the dimension of length. In particular, 
if $r_0=a\ln\left[\f{\a+\b}{\a-\b}\right]$, with $\a>\b$, then the potential has a minimum value of $-\f{(\a-\b)^2}{4\b}$.
By construction, $f_4$ approximation is more effective around $r=r_0$ region \cite{pekeris.ap3,Pekeris,pekeris.badawi,pekeris.ap4}. 
In this article, we propose a new approximation to the centrifugal term, following our previous works 
\cite{nath2021,nath2021.epjp}, 
\beq\label{app.convex}
\f{1}{r^2}\approx f_5(r)=\sum\limits_{j=1}^4\lam_j\,f_{j}(r)=\f{1}{a^2}\left(y_1+\f{y_2s}{1-s}+\f{y_3s^2}{(1-s)^2}\right),
\eeq 
where
\beq
y_i=\sum\limits_{j=1}^4\lam_jx_{ij},~i=1,2,3; \ \ \ \ \ \ \ \ \ \ \ \ \
\sum\limits_{j=1}^{4}\lam_j=1.
\eeq
Figure~\ref{Fig1.approx} compares these approximations with \emph{exact} centrifugal term $\f{\ell(\ell+1)}{r^2}$, for 
$\ell=2$ having following parameters: $(\hbar,\bar{\mu})=(1,1)$, $(a,\a,\b)=(1/0.025,1/a,0.0001)$, $(\xi_1,\xi_2)=(1.1,0.98)$, $D=3$. 
The upper and lower panels correspond to $r \to 0$ and $r \to r_0$ regions. In each row, five approximations are illustrated in 
five panels--the first four refer to $f_1, f_2, f_3, f_4$, while two rightmost panels imply $f_5^{(a)}$ and $f_5^{(b)}$ 
having $(\lam_1,\lam_2,\lam_3,\lam_4)=(0,0,0.98,0.02)$ and $(0,0,0.02,0.98)$ respectively. Clearly, panels (C), (E) show that $f_3$ and $f_5^{(a)}$ are good approximations near origin. Likewise,  
from panels (I), (J), it is inferred that $f_4,f_5^{(b)}$ are superior in the neighborhood of $r=r_0$. 
Thus $f_5$ remains quite accurate over the entire effective domain of $r$ \cite{nath2021,nath2021.epjp}, which has been defined 
in our previous communications \cite{effective.domain,effective.domain2}. Throughout the article, for sake of convenience, $f_5$ approximations 
will be used for various sets of $\{\lam_1, \lam_2, \lam_3, \lam_4\}$. These will be indicated by alphabetic superscripts. 

\subsection{Ro-vibrational energy by NU method}
{\color{red} This is a well-established mathematical tool for various eigenvalue problems \cite{NU}. It has been utilized by a number of 
researchers for a broad range central and non-central potentials in non-relativistic and relativistic quantum mechanics \cite{ahmadov2014, 
ahmadov2017, ahmadov2019, ahmadov2020}.} Under the transformation, $s=e^{-r/a}$, 
Eq.(\ref{Eq.R}) becomes,
\beq
\ba{l}\label{Eq.R.NU.Ec}
\ds\f{d^2R}{ds^2}+\f{\widetilde{\tau}(s)}{\sigma(s)}\f{dR}{ds}+\f{\widetilde{\sigma}(s)}{\left[\sigma(s)\right]^2}R=0,
\ea 
\eeq 
where
\beq
\ba{ll}
\widetilde{\tau}(s)&=1-s,~ \sigma(s)=s-s^2,~ \widetilde{\sigma}(s)=-As^2+Bs-C,
\\
A &
=\f{2\bar{\mu}a^2\a}{\hbar^2}-L(L+1)(y_2-y_3)+C,\\
B &
=\f{2\bar{\mu}a^2(\a-\b)}{\hbar^2}-L(L+1)y_2+2C,\\
C &
=\epsilon_0^2+L(L+1)y_1,\\
\epsilon_0^2&=-\f{2\bar{\mu}a^2 E}{\hbar^2}.
\ea
\eeq 
Let us suppose, 
\beq
R(s)=\phi_1(s)\phi_2(s),
\eeq
be the solution of Eq.~(\ref{Eq.R.NU.Ec}). Following \cite{nath2021,nath2021.epjp,NU}, it is then obtained that, 
\beq\label{Eq.ys}
\sigma(s)\phi_2^{''}(s)+\tau(s)\phi_2'(s)+\nu\, \phi_2(s)=0,
\eeq
and
\beq
\ba{l}
\phi_1(s)=\ds e^{\int\f{\pi(s)}{\sigma(s)}ds},
\ea
\eeq 
where
\beq
\ba{ll}\label{pi.sigmas}
\pi(s)&=\f{\sigma'(s)-\widetilde{\tau}(s)}{2}\pm\sqrt{\left(\f{\sigma'(s)-\widetilde{\tau}(s)}{2}\right)^2-\widetilde{\sigma}(s)+k\,\sigma(s)},\\
\tau(s)&=\widetilde{\tau}(s)+2\pi(s),~
\bar{\nu}=k+\pi'(s),
\ea 
\eeq
with $\bar{\nu}$ and $k$ being real constants. 
Since $\pi(s)$ is a polynomial in $s$, one has to find $k$ in such a way that, 
$\left(\f{\sigma'(s)-\widetilde{\tau}(s)}{2}\right)^2-\widetilde{\sigma}(s)+k\,\sigma(s)$ is the square of a polynomial 
in $s$. Then the solutions of Eq.~(\ref{Eq.ys}) can be written as follows,  
\beq
\phi_{2,n_r}(s)=\f{1}{\rho(s)}\f{d^{n_r}}{ds^{n_r}}\left[\sigma^{n_r}(s)\rho(s)\right],
\eeq
and the corresponding eigenvalues are obtained as, 
\beq\label{eigen.nu}
\nu_{n_r}=-{n_r}\,\tau'(s)-\f{{n_r}({n_r}-1)}{2}\sigma^{''}(s),~{n_r}=0,1,2,\dots, 
\eeq 
where
\beq
\ba{l}
\rho(s)=\ds \left[\sigma(s)\right]^{-1}\,e^{\int\f{\tau(s)}{\sigma(s)}ds}.
\ea 
\eeq  

According to NU method, and following the prescription \cite{nath2021,nath2021.epjp,NU}, one gets a pair of $k$ as, 
\beq
\ba{ll}\label{kpitaunu.Ec}
k_{\pm}
=\f{2\bar{\mu}a^2(\a-\b)}{\hbar^2}-L(L+1)y_2\pm\sqrt{C(2L_1-1)},
\ea
\eeq
where
\beq
L_1
=\f{1}{2}+\sqrt{\f{1}{4}+\f{2\bar{\mu}a^2\b}{\hbar^2}+L(L+1)y_3}.
\eeq 
Now, application of Eq.~(\ref{kpitaunu.Ec}) in Eq.~(\ref{pi.sigmas}) yields,   
\beq
\ba{ll}\label{pi.tau.nu.Ec}
\pi(s)
&=-\f{s}{2}\pm\left\{\ba{lll}\left(\sqrt{C}-L_1+\f{1}{2}\right)s+\sqrt{C},&k=k_+,& k_+-B>0\\\left(\sqrt{C}+L_1-\f{1}{2}\right)s+\sqrt{C},&k=k_-,& k_--B>0\\
\left(\sqrt{C}-L_1+\f{1}{2}\right)s-\sqrt{C},&k=k_+,& k_+-B<0\\\left(\sqrt{C}+L_1-\f{1}{2}\right)s-\sqrt{C},&k=k_-,& k_--B<0\ea \right\}.
\ea
\eeq
We have chosen $k=k_-$ with $k_--B<0$, and selected, 
\beq\label{rho.phi2}
\pi(s)=-\left(\sqrt{C}+L_1\right)s+\sqrt{C}.
\eeq
This leads to \cite{Gradshteyn},     
\beq
\ba{l}
\rho(s)=s^{2\sqrt{C}}(1-s)^{2L_1-1},\\
\phi_1(s)=s^{\sqrt{C}}(1-s)^{L_1},
\ea 
\eeq
and
\beq
\ba{ll}
\phi_{2,n_r}&=({n_r})!\,P_{n_r}^{(2\sqrt{C},2L_1-1)}(1-2s),\\
&=\f{\G\left(n_r+2\sqrt{C}+1\right)}{\G\left(2\sqrt{C}+1\right)}{}_2F_1\left(-n_r,n_r+2\sqrt{C}+2L_1;2\sqrt{C}+1;s\right),
\ea
\eeq 
where $P_n^{(a,b)}(x)$ is the Jacobi polynomial in $x$ of degree $n$ with parameters $a,b$, while ${}_2F_1(a,b;c;x)$ is the 
Hypergeometric function, defined by ${}_2F_1(a,b;c;x)=\sum\limits_{k=0}^{\infty}\f{(a)_k(b)_k}{(c)_k}\f{x^k}{k!}$ and 
$(a)_k=\f{\G(a+k)}{\G(a)}=a(a+1)(a+2)...(a+k-1)$ is the Pochhammer symbol.
The eigenvalues $E_{n_r,l}$ are then given as, 
\beq\label{Energy.MR}
\ba{ll}
E_{n_r,\ell}
&=\f{\hbar^2L(L+1) y_1}{2\bar{\mu}a^2}-\f{\hbar^2}{2\bar{\mu}a^2}\left(\f{\f{\bar{\mu}a^2\a}{\hbar^2}-L(L+1)(y_2-y_3)/2}{n_r+L_1}-\f{n_r+L_1}{2}\right)^2.
\ea 
\eeq
Therefore, the radial wave function can be expressed as \cite{Gradshteyn}, 
\beq
R_{n_r,\ell}(r)=N_{{n_r},\ell}\,s^{\sqrt{C_{n_r,\ell}}}(1-s)^{L_1}\,{}_2F_1\left(-{n_r},{n_r}+2\sqrt{C_{n_r,\ell}}+2L_1;2\sqrt{C_{n_r,\ell}}+1;s\right),
\eeq 
where 
\beq
N_{{n_r},\ell}=\left[\f{2\sqrt{C_{n_r,\ell}}(n_r+\sqrt{C_{n_r,\ell}}+L_1)\G(n_r+2\sqrt{C_{n_r,\ell}}+1)\G(n_r+2\sqrt{C_{n_r,\ell}}+2L_1)}{a({n_r})!(n_r+\sqrt{C_{n_r,\ell}})\G(n_r+2L_1)\left[\G(2\sqrt{C_{n_r,\ell}}+1)\right]^2}\right]^{\f{1}{2}},
\eeq 
is the normalization constant, and $C_{n_r,\ell}=-\f{2\bar{\mu}a^2E_{n_r,\ell}}{\hbar^2}+(\ell+\f{D-3}{2})(\ell+\f{D-1}{2})x_1$. Finally, the explicit form of eigenfunctions can be written as, 
\beq
\ba{lr}
\psi_{n_r,\ell,\{\mu\}}({\bf r})&=N_{{n_r},\ell}r^{\f{1-D}{2}}\,s^{\sqrt{C_{n_r,\ell}}}(1-s)^{L_1}\,{}_2F_1\left(-{n_r},{n_r}+2\sqrt{C_{n_r,\ell}}+2L_1;2\sqrt{C_{n_r,\ell}}+1;s\right)Y_{\ell,\{\mu\}}(\Om),\\
&s=e^{-r/a}.
\ea
\eeq
It is to be noted that the approximation (\ref{app.convex}) is well defined for convex combination $(0\le \lam_j\le 1,~j=1(1)4,~\sum\limits_{j=1}^4\lam_j=1)$ for Eckart potential, and one may consider a linear combination (may assume negative 
values) for different values of $\lam_j,~j=1(1)4$, if they satisfy the following relations, 
\beq
\ba{l}\label{relation1.Ec}
\sum\limits_{j=1}^4\lam_j=1,\\
\sum\limits_{j=1}^4\lam_jx_{1j}\ge 0,\\
\f{1}{4}+\f{2\bar{\mu}a^2\b}{\hbar^2}+(\ell+\f{D-3}{2})(\ell+\f{D-1}{2})\sum\limits_{j=1}^4\lam_jx_{3j}\ge 0.
\ea
\eeq

\begingroup            
\squeezetable
\begin{table}
	\caption{\label{Table1} Energies, $E_{n_r,\ell}$, within different approximations, for $\hbar=\bar{\mu}=1,\a=\f{1}{a},a=1/0.025,D=3$.
The upper and lower portions refer to $\b=0.0001$ and 0.0005. For details, see text.}
	\centering
        \begin{ruledtabular}
	\begin{tabular}{cccccccccc}
		$n_r$	& $\ell$ & $-E_{n_r,\ell}[f_1]$\footnotemark[1]&$-E_{n_r,\ell}[f_2]$\footnotemark[2]	& $-E_{n_r,\ell}[f_3]$ & $-E_{n_r,\ell}[f_4]$ & $-E_{n_r,\ell}[f_5^{(c)}]$ & $-E_{n_r,\ell}[f_5^{(d)}]$& Ref.~\cite{lucha1999} & GPS\footnotemark[3] \\ \hline
\multicolumn{10}{c}{$\b=0.0001$} \\
\hline
		0  &  1   &  0.1008879   &  0.1015944  &  0.1008358  &  0.1008358  &  0.1010119  &  0.1008410  &0.1008358   & 0.1008359\\
		0  &  2   &  0.0415198   &  0.0414791  &  0.0413635  &  0.0413635  &  0.0414643  &  0.0413792  & 0.0413642  & 0.0413642\\
 		0  &  3   &  0.0193308   &  0.0189461  &  0.0190183  &  0.0190183  &  0.0191600  &  0.0190495  &0.0190220   & 0.0190220\\
		1  &  1   &  0.0401768   &  0.0403163  &  0.0401247  &  0.0401247  &  0.0401887  &  0.0401299  & 0.0401250  & 0.0401250\\
		1  &  2   &  0.0190752   &  0.0189774  &  0.0189190  &  0.0189190 &  0.0190087  &  0.0189346  & 0.0189216  & 0.0189216\\
 		1  &  3   &  0.0091303   &  0.0088858  &  0.0088178  &  0.0088178  &  0.0089875  &  0.0088491  & 0.0088297  & 0.0088297\\
		2  &  1   &  0.0185142   &  0.0185468  &  0.0184622  &  0.0184622  &  0.0185050  &  0.0184674  & 0.0184632  & 0.0184632\\
		2  &  2   &  0.0090066   &  0.0089303  &  0.0088504  &  0.0088504  &  0.0089444  &  0.0088660  & 0.0088576  & 0.0088576\\
 		2  &  3   &  0.0040362   &  0.0038908  &  0.0037237  &  0.0037237  &  0.0039132  &  0.0037550  & 0.0037525  & 0.0037525\\ \hline
\multicolumn{10}{c}{$\b=0.0005$} \\
\hline
		0  &  1   &  0.0704527   &  0.0706770  &  0.0704006  &  0.0704006  &  0.0704815  &  0.0704058  & $\cdots$   & 0.0704007 \\
		0  &  2   &  0.0342158   &  0.0340877  &  0.0340595  &  0.0340596  &  0.0341431  &  0.0340752  & $\cdots$ & 0.0340604\\
 		0  &  3   &  0.0169482   &  0.0165709  &  0.0166357  &  0.0166359  &  0.0167790  &  0.0166671  & $\cdots$ & 0.0166401\\
		1  &  1   &  0.0301503   &  0.0301782  &  0.0300982  &  0.0300982  &  0.0301402  &  0.0301034  & $\cdots$ & 0.0300987 \\
		1  &  2   &  0.0159649   &  0.0158461  &  0.0158087  &  0.0158088  &  0.0158942  &  0.0158244  & $\cdots$ & 0.0158120 \\
 		1  &  3   &  0.0079688   &  0.0077346  &  0.0076563  &  0.0076564  &  0.0078281  &  0.0076877  & $\cdots$ & 0.0076699 \\
		2  &  1   &  0.0141676   &  0.0141623  &  0.0141156  &  0.0141156  &  0.0141509  &  0.0141208  & $\cdots$ & 0.0141170 \\
		2  &  2   &  0.0074840   &  0.0074025  &  0.0073277  &  0.0073278  &  0.0074208  &  0.0073434  & $\cdots$ & 0.0073363 \\
 		2  &  3   &  0.0034413   &  0.0033048  &  0.0031288  &  0.0031290  &  0.0033201  &  0.0031602  & $\cdots$ & 0.0031612 \\
	\end{tabular}
\end{ruledtabular}
	\begin{tabbing}
		\footnotemark[1]{These compare with those from \cite{dong2007}.} \hspace{0.8in} \=
		\footnotemark[2]{These compare with those from \cite{taskin2010}, where $\xi=1.10$ and $\lam=0.98$.}\\
		\footnotemark[3]{These are calculated here for this work, using GPS \cite{gps1, gps2, gps3}.}
	\end{tabbing} 
\end{table}	

\section{Results and discussion}\label{Sec3.results}
At first, Table~\ref{Table1} reports 9 low-lying energies corresponding to first three non-zero-$\ell$ (1-3) values of first three $n_r$ (1--3). These are given for $D=3$ having a fixed $a=1/0.025$; the upper and lower portions correspond to $\b=0.0001$ and 0.0005 respectively. The four columns 3--6 employ four different approximations \emph{viz.}, $f_1, f_2, f_3, f_4$ from Eqs.~(\ref{greene.app1}) and (\ref{pekeris.app1}), while columns 7 and 8 provide those for approximations $f_5^{(c)}$ and $f_5^{(d)}$ in Eq.~(\ref{app.convex}), with fixed $(\xi_1,\xi_2)=(1.1,0.98)$ set, but having different $\lam_i$'s, namely, $(\lam_1,\lam_2,\lam_3,\lam_4)=(0.5, 0.2, 0.2, 0.1)$ and $(0.1, 0, 0, 0.9)$. These are compared with accurate numerical results of \cite{dong2007,taskin2010,lucha1999} and from the generalized pseudospectral (GPS) method \cite{gps1, gps2, gps3}. The latter has been found to be very successful for a variety of potentials of physical and chemical relevance. In case of $\beta=0.0005$, no reference energies could be found except the GPS \cite{gps1, gps2, gps3}. The approximated energies quite favorably compare with literature results. 



In our analysis, for $D=3$, under $f_1$ approximation, the zero-energy states characterized by quantum numbers $(n_r,\ell)$, for a particular $a=a_0$, 
can be obtained from the following equation, 
\beq
\ba{l}\label{a4th}
\ds\left[\f{8\bar{\mu}a_0^2(\a_0-\b)}{\hbar^2}-\left(2n_r+1\right)^2-\left(2\ell+1\right)^2\right]^2=\ds (2n_r+1)^2\left[(2\ell+1)^2+\f{8\bar{\mu}a_0^2\b}{\hbar^2}\right], 
\ea 
\eeq 
where $\a_0$ is a function of $a_0$. 
Our analysis also reveals that, for a given $a=a_{12}$, two states identified by quantum numbers $({n_{r1},\ell_1})$ and $({n_{r2},\ell_2})$ are 
degenerate possessing same energy. This can be found from following relation,  
\beq
\ba{l}\label{degenerate.D3}
\ds\f{2\bar{\mu}a_{12}^2\a_{12}}{\hbar^2}\left(\f{1}{n_{r1}+\f{1}{2}+\sqrt{\ell_1(\ell_1+1)+\f{1}{4}+\f{2\bar{\mu}a_{12}^2\b}{\hbar^2}}}\mp\f{1}{n_{r2}+\f{1}{2}+\sqrt{\ell_2(\ell_2+1)+\f{1}{4}+\f{2\bar{\mu}a_{12}^2\b}{\hbar^2}}}\right)\\\\
\ds=\f{1\mp1}{2}+n_{r1}\mp n_{r2}+\sqrt{\ell_1(\ell_1+1)+\f{1}{4}+\f{2\bar{\mu}a_{12}^2\b}{\hbar^2}}\mp\sqrt{\ell_2(\ell_2+1)+\f{1}{4}+\f{2\bar{\mu}a_{12}^2\b}{\hbar^2}}.
\ea 
\eeq 
Considering the positive sign, $a_{12}$ seems to satisfy the equation as below, 
\beq
\ba{l}\label{degenerate.D32}
\ds\f{2\bar{\mu}a_{12}^2\a_{12}}{\hbar^2}
\ds=\left(n_{r1}+\f{1}{2}+\sqrt{\left(\ell_1+\f{1}{2}\right)^2+\f{2\bar{\mu}a_{12}^2\b}{\hbar^2}}\right)\left(n_{r2}+\f{1}{2}+\sqrt{\left(\ell_2+\f{1}{2}\right)^2+\f{2\bar{\mu}a_{12}^2\b}{\hbar^2}}\right).
\ea 
\eeq
The degenerate energy levels can be obtained numerically under $f_5$ approximation. Using Eq.~(\ref{degenerate.D32}), one can find these levels, $E_{n_{r1},2}=E_{n_{r2},2}$ which are equal at the points, $a=a_{12}$. 

\begingroup
\squeezetable
\begin{table}
	\caption{\label{Table5} Energies, $E_{n_r,\ell}$ for $\hbar=\bar{\mu}=1,\a=1/a,\b=0.0001,a=1/0.025, (\xi_1,\xi_2)=(1.1,0.98)$, within $f_5^{(d)}$
approximation. The columns (3,8), (4,9), (5,10) refer to $D=3,4,5$. See text for details.}
	\centering
	\begin{ruledtabular}
	\begin{tabular}{ccccc|ccccc} 
		$n_r$ & $\ell$ &  $-E_{n_r,\ell}(3)$ &  $-E_{n_r,\ell}(4)$ & $-E_{n_r,\ell}(5)$ &$n_r$ & $\ell$ &  $-E_{n_r,\ell}(3)$ & $-E_{n_r,\ell}(4)$ & $-E_{n_r,\ell}(5)$ \\ \hline
		0 & 1& 0.1008410  & 0.0631369   & 0.0413635  &1 & 3& 0.0088491  & 0.0058207    & 0.0035513  \\
		
		0 & 2& 0.0413792  & 0.0278917   & 0.0190183  &1 & 4 & 0.0036034 & 0.0019722    & 0.0006987   \\
		
		0 & 3& 0.0190495  & 0.0130013   & 0.0086797   &2 & 1& 0.0184674  & 0.0129021  & 0.0088504  \\
		
		0 & 4 & 0.0087318 & 0.00564956   & 0.0033141   &2 & 2& 0.0088660  & 0.0059195    & 0.0037237  \\
		
		1 & 1& 0.0401299  & 0.0274586   & 0.0189190  & 2 & 3& 0.0037550  & 0.0021629    & 0.0009477   \\
		
		1 & 2& 0.0189346  & 0.0130356   & 0.0088178   &2 & 4& 0.0009998  & 0.0001657    & $\cdots$\\
	\end{tabular}
	\end{ruledtabular}
\end{table}	
In Table~\ref{Table5}, we offer representative state energies, for different dimensions (3--5) for particular values of $a=1/0.025, \ \beta=0.0001$, 
within the $f_5^{(d)}$ approximation, having $\lam_1-\lam_4$ as $(0.1,0,0,0.9)$. The $D=3$ results are slightly different from those in previous
table, as the $\lam_i$'s employed for $f_5$ are different. No reference energies can be found for comparison. Note that, the state $E_{2,4}$ does not 
exist for $D=5$, for the $a$ considered in this table.  
Furthermore, under $f_1$ approximation, for $D\ge 3$, there occurs an inter-dimensional degeneracy, corresponding to energy levels $E_{n_{r1},\ell_1}(D_{i})=E_{n_{r2},\ell_2}(D_{j})$ at the point $a=a_{ij}$, given by the equation, 
\beq
\ba{l}\label{degenerate.gD}
\ds\f {2\bar{\mu} \left(a_{12}^{D_{ij}}\right)^2 \a_{12}^{D_{ij}}} {\hbar^2} 
\left(\f{1}{n_{r1}+\f{1}{2}+\sqrt{\Big(\ell_1+\f{D_{i}-2}{2}\Big)^2++\f{2\bar{\mu}\left(a_{12}^{D_{ij}}\right)^2\b}{\hbar^2}}}\mp\f{1}{n_{r2}+\f{1}{2}+
\sqrt{\Big(\ell_2+\f{D_{j}-2}{2}\Big)^2+\f{2\bar{\mu}\left(a_{12}^{D_{ij}}\right)^2\b}{\hbar^2}}}\right)\\\\
\ds=\f{1\mp1}{2}+n_{r1}\mp n_{r2}+\sqrt{\Big(\ell_1+\f{D_{i}-2}{2}\Big)^2+
\f{2\bar{\mu}\left(a_{12}^{D_{ij}}\right)^2\b}{\hbar^2}}\mp\sqrt{\Big(\ell_2+\f{D_{j}-2}{2}\Big)^2+\f{2\bar{\mu}\left(a_{12}^{D_{ij}}\right)^2\b}{\hbar^2}}
\ea 
\eeq 
where $\a_{12}^{D_{ij}}$ is again a function of $a_{12}^{D_{ij}}$. Similarly, considering the positive sign, one obtains, 
\beq
\ba{l}\label{degenerate.gD2}
\ds\f{2\bar{\mu}\left(a_{12}^{D_{ij}}\right)^2\a_{12}^{D_{ij}}}{\hbar^2}=
\ds\left(\f{1}{2}+n_{r1}+\sqrt{\left(\ell_1+\f{D_{i}-2}{2}\right)^2+\f{2\bar{\mu}\left(a_{12}^{D_{ij}}\right)^2\b}{\hbar^2}}\right) \\
\left(\f{1}{2}+n_{r2}+\sqrt{\left(\ell_2+\f{D_{j}-2}{2}\right)^2+\f{2\bar{\mu}\left(a_{12}^{D_{ij}}\right)^2\b}{\hbar^2}}\right).
\ea 
\eeq
The value of $a=a_{0}^{D}$, corresponding to a zero-energy state occurs for $(n_r,\ell)$ quantum numbers, which for $D>3$, under $f_1$ 
approximation, can be estimated from, 
\beq
\ba{l}\label{a4th2}
\ds\left[\f{8\bar{\mu}(a_{0}^{D})^2(\a_{0}^{D}-\b)}{\hbar^2}-\left(2n_r+1\right)^2-\left(2\ell+D-2\right)^2\right]^2=
\ds (2n_r+1)^2\left[\left(2\ell+D-2\right)^2+\f{8\bar{\mu}(a_{0}^{D})^2\b}{\hbar^2}\right].
\ea 
\eeq
From Eqs.~(\ref{degenerate.gD2}) and (\ref{a4th2}) we see that, $a_{12}^{D_{ij}}=a_{12},$ if $D_i=D_j=3$ and $a_{0}^{D}=a_0$ if $D=3$. 


\section{Conclusion}\label{Sec4.con}
Analytical eigenvalues and eigenfunctions are obtained for {\color{red} Eckart potential from a solution of $D$-dimensional SE} within the rubric of Nikiforov-Uvarov method. A new, improved approximation to the centrifugal term is proposed for this potential, which is based on physical grounds. 
{\color{red} This general form is intuitively arrived from an amalgamation of two well-known existing approximations for centrifugal potential, namely, 
Greene-Aldrich and Pekeris-type. This required a linear combination of four approximating functions. The condition on the involved parameters is also 
discussed.} It offers an accurate representation of the potential throughout the entire effective domain of $r$. Energy states corresponding to arbitrary quantum numbers ($\ell \neq 0$) are presented and compared critically amongst each other, as well as, with available theoretical reference results. 
{\color{red} The approach can be easily extended to other exponential-type potentials of interest, such as Deng-Fan, Manning-Rosen, P\"oschl-Teller, 
Hulth\'en etc. Some of these may be taken up in future communications.} 

\section*{Acknowledgement}
Financial support from MATRICS, DST-SERB, New Delhi (sanction order: MTR/2019/000012) is gratefully acknowledged. Partial funding from DST SERB 
(sanction order: CRG/2019/000293) is appreciated.  
\section*{ORCID}
\noindent Debraj Nath: https://orcid.org/0000-0001-9937-7032\\
\noindent Amlan K. Roy: https://orcid.org/0000-0001-5555-8915

\end{document}